\documentclass[12pt]{article}
\usepackage{amsmath,amssymb,amsfonts,epsfig}
\usepackage{graphics}
\usepackage{amssymb,amsmath}
\usepackage{cite}
\usepackage[usenames]{color}
\usepackage[usenames,dvipsnames]{pstricks}
\usepackage{epsfig}
\usepackage{pst-grad} 
\usepackage{pst-plot} 
\newcommand{\ct}{\cite}
\newcommand{\lb}{\label}

\newcommand{\bc}{\begin{center}}
\newcommand{\ec}{\end{center}}
\newcommand{\bd}{\begin{displaymath}}
\newcommand{\ed}{\end{displaymath}}
\newcommand{\be}{\begin{equation}}
\newcommand{\ee}{\end{equation}}
\newcommand{\ba}{\begin{array}}
\newcommand{\ea}{\end{array}}
\newcommand{\bea}{\begin{eqnarray}}
\newcommand{\eea}{\end{eqnarray}}
\newcommand{\bt}{\begin{tabular}}
\newcommand{\et}{\end{tabular}}

\newcommand{\bp}{\begin{picture}}
\newcommand{\ep}{\end{picture}}
\newcommand{\bfi}{\begin{figure}}
\newcommand{\efi}{\end{figure}}

\begin{document}

\begin{titlepage}

\begin{center}

{\bf Dedicated to Holger Bech Nielsen on his 70th birthday.}
\end{center}

\vspace*{0.5cm}

\begin{center}
{{\Large {\bf Quantum Gravity in Plebanski Formulation} }}
\vspace*{10mm}

{\bf \large{ L. V. Laperashvili }}
\end{center}
\begin{center}
\vspace*{0.4cm} {\it {The Institute of Theoretical and Experimental Physics,\\
Bolshaya Cheremushkinskaya, 25, 117218 Moscow, Russia}}\\
\vskip 0.2cm \centerline{\tt laper@itep.ru}

\vspace*{1.0cm}

\end{center}

\begin{center}{\bf Abstract}\end{center}

\begin{quote}

We present a theory of four-dimensional quantum gravity with
massive gravitons which may be essentially renormalizable. In the
Plebanski formulation of General Relativity (GR), in which the
tetrads, the connection and the curvature are all independent
variables (and the usual relations among these quantities are
valid only on-shell), we consider a nonperturbative theory of
gravity with a nonzero background connection. We predict a tiny
value of the graviton mass: $m_g\approx 1.8\times
10^{-42}\,{\mbox{GeV}}$ and an extremely small dimensionless
coupling constant of the perturbative gravitational interaction:
$g\sim 10^{-61}$. We put forward the idea by H.~Isimori \ct{15} on
renormalizability of quantum gravity having multi-gravitons with
masses $m_0, m_1,..., m_{N-1}$.

\end{quote}

\end{titlepage}

\section{Introduction}

Quantum gravity aims to solve the problem of merging quantum
mechanics and general relativity, the two great conceptual
revolutions in twentieth century physics. Gravity is the only
fundamental interaction which cannot be considered as a full
quantum theory. However, the theory of gravity is not complete
yet. Currently, it cannot answer the following profound questions:
What is the microstructure of spacetime explaining macroscopic
gravitational interactions?  What are the quantum origins of
space, time and our Universe? Are "space"\,, "time"\, and
"causality"\, fundamental concepts? Where is the space-time
geometry allowed to undergo large quantum fluctuations?

A lot of articles in literature are devoted to quantum gravity. We
would like to mention the following books
\ct{1a,1b,1c,1d,1e,1f,1g,1h,1i,1j,1k,1l,1m,1n,1p,1q}, reviews
\ct{2a,2b,2c} and articles dealing with possible violations of
Lorentz invariance at high energies
\ct{3a,3b,3c,3d,3dd,3e,3f,3ff,3g,3gg,3h,3i}. The most important
question in quantum gravity is whether the ground state of
space-time obeys Lorentz invariance. The violation of Lorentz
invariance appears in many approaches to quantum gravity. In
addition, Lorentz-violating models also appear in  noncommutative
geometry \ct{4a,4b}.

It is well-known that quantum gravity is a nonrenormalizable
theory. However, in the present paper we investigate the
four-dimensional quantum gravity which may be essentially
renormalizable if one relaxes the assumption of metricity of the
theory. Here we consider the Plebanski formulation \ct{1} of
general relativity in which the tetrads, the connection and the
curvature are all independent dynamical variables, and the
relations existing among them are valid only on-shell.

\section{Plebanski theory of gravity}

It has been showed \ct{1,2,3,4,5} that the first-order formulation
of gravity that uses tetrads instead of the metric is more
fundamental. In the Plebanski formulation of theory of gravity
\ct{1} there arises a new independent field -- the connection
$A^{IJ}$, and the tetrad $\theta^I$ is used instead of the metric
$g_{\mu\nu}$. Both $A^{IJ}$ and $\theta^I$ are one-forms. Indices
$I,J = 0,1,2,3$ refer to the flat space-time with Minkowski metric
$\eta_{IJ}$: $\eta^{IJ} = {\rm diag}(-1, 1, 1, 1)$. This is a flat
space which is tangential to the curved space with metric
$g_{\mu\nu}$ at each its point. The world interval is represented
as $ds^2 =  \eta_{IJ}\theta^I \otimes \theta^J$, i.e. $g_{\mu\nu}
= \eta_{IJ} \theta^I_{\mu}\otimes \theta^J_{\nu}$.

The action of the first-order gravity is \ct{1,2,3}:
\be S(\theta, A) = \frac{1}{\kappa^2}\int_M \epsilon^{IJKL} \left(
\theta^I \theta^J F^{KL} + \frac{\Lambda}{4}\theta^I \theta^J
\theta^K \theta^L \right).    \lb{1} \ee
Here $\kappa^2=8\pi G$, where $G$ is the gravitational coupling
constant; $F = dA+(1/2)[A,A]$ is the curvature of the Lorentz
group Lie algebra with spin connection A, and $\Lambda$ is (a
multiple of) the cosmological constant. The wedge product of all
the forms is assumed.

The next step is to construct two-forms $\theta^I \wedge \theta^J$
and take their self-dual parts with respect to the indices $I,J$.
Then we introduce an arbitrary time plus space split of the
internal indices $I = (0, a),\,\, a = 1, 2, 3$, and construct $F^a
= \frac {1}{2}F^a_{\mu\nu}dx^{\mu}\wedge dx^{\nu}$ which is the
field-strength two-form for gauge connection $A^a =
A^a_{\mu}dx^{\mu}$. The field strength is written in component
form as $F^a_{\mu\nu} =
\partial_{\mu}A^a_{\nu} -
\partial_{\nu}A^a_{\mu} + f^{abc}A^b_{\mu}A^c_{\nu}$, with $SO(3,C)$ structure
constants $f^{abc} = \epsilon^{abc}$.

We try to get some new insight into the renormalization properties
of 3+1 gravity by using the above action as a starting point.
First, we see a problem that there is no quadratic term in
(\ref{1})  that could be interpreted as kinetic. Such a term
arises only if one assumes the background to be constant (see
below).

The main idea of Ref.~\ct{1} is that, in addition to the
tetrad-like and connection fields, the theory contains a new field
which on-shell becomes identified with the Weyl part of the
curvature tensor. In this formulation gravity becomes a non-metric
theory: instead of the tetrad one-forms one uses certain new
two-forms that become related to the metric only on-shell.

In this paper we consider the original self-dual formulation of
Plebanski \ct{1,2,3}. In the units $\kappa = 1$, we adapt the
starting action (\ref{1}) to the language of the SO(3,C) gauge
algebra as
\be
    S(\Sigma,A,\psi) = \int_{M_4} \left( \Sigma^a \wedge F^a +
    (\Psi^{-1})_{ab}
    \Sigma^a  \wedge \Sigma^b \right),    \lb{2} \ee
where $\Sigma^a = \frac {1}{2} \Sigma^a_{\mu\nu}dx^{\mu} \wedge
dx^{\nu}$ is a triplet of the SO(3,C) two-forms:
\be  \Sigma^a = i \theta^0 \wedge  \theta^a -\frac 12
\epsilon^{abc}\theta^b \wedge \theta^c.  \lb{3} \ee
Here $a, b, c,...=1,2,3$ are the SU(2) Lie algebra indices. In
Eq.~(\ref{2}) we have \ct{2,3}:
\be (\Psi^{-1})_{ab} = \Lambda \delta_{ab} + \psi_{ab}, \lb{4} \ee
where $\psi_{ab}$ is a field that on-shell becomes the Weyl part
of the curvature, symmetric and traceless.

The equations of motion resulting from (\ref{2}) are (see \ct{3}):
\be  \frac{\delta S}{\delta A^a} = D\Sigma^a = d\Sigma^a +
\epsilon^a_{bc}A^b \wedge \Sigma^c = 0,    \lb{5} \ee
\be  \frac{\delta S}{\delta \psi_{ab}} = \Sigma^a \wedge \Sigma^b
- \frac 13 \delta^{ab} \Sigma_c \wedge \Sigma^c = 0,    \lb{6} \ee
\be     \frac{\delta S}{\delta \Sigma^a} = F^a -
(\Psi^{-1})^a_b\Sigma^b = 0.     \lb{7} \ee
The Eq.~(\ref{5}) states that $A^a$ is the self-dual part of the
spin connection compatible with the two-forms $\Sigma^a$, where
$D$ is the exterior covariant derivative with respect to $A^a$.
Eq.~(\ref{6}) implies that the two-forms $\Sigma^a$ can be
constructed from tetrad one-forms giving (\ref{3}), which fixes
the conformal class of the space-time metric $g_{\mu\nu} =
\eta_{IJ}\theta^I_{\mu} \otimes \theta^J_{\nu}$ defined by the
tetrads. The Eq.~(\ref{7}) states that the curvature $F^a$ is
self-dual as a two-form, which implies that the metric
$g_{\mu\nu}$ derived from the tetrad one-forms $\theta^I$
satisfies the vacuum Einstein equations.

The 2-form fields $\Sigma^a$ can therefore be integrated out of
Eq.~(\ref{2}). Thus, we are led to Einstein's gravity given by the
form:
\be
    S(A,\psi) = \int_M {(\delta^{ab}\Lambda +\psi^{ab})}^{-1}F^a\wedge F^b,  \lb{8} \ee
discussed in Refs.~\ct{2,3,4,5}.

\section{Nonperturbative theory of Plebanski gravity}

Here we present a version of the nonperturbative gravity, assuming
the condition $\Lambda \gg |\psi|$, i.e. considering $\psi_{ab}$
as a perturbation to $\Lambda \delta_{ab}$, and use the following
expansion:
\be {(\delta^{ab}\Lambda +\psi^{ab})}^{-1}F^a\wedge F^b\approx
\frac{1}{\Lambda}(\delta_{ab} - \frac{1}{\Lambda}\psi_{ab} +
\frac{1}{\Lambda^2}\psi_{ac}\psi_{cb} +...)F^a\wedge F^b.
\lb{9} \ee
According to Ref.~\ct{2}, in the action (\ref{2}) we have:
\be \Sigma^a  \wedge \Sigma^a = -6i\sqrt{-g}d^4x,   \lb{9a} \ee
what means that the dimensionless cosmological constant is equal
to
 \be \Lambda_0 = \frac{\rho_{vac}}{(M_{Pl}^{red.})^4} = 6\Lambda, \lb{10} \ee
where $\rho_{vac}$ is the effective vacuum energy density of the
Universe and $M_{Pl}^{red.}=1/{\sqrt{8\pi G}}$ is the reduced
Planck mass.

Now we can calculate the partition function in Euclidean space:
\be    Z = \int[{\cal D}A ][{\cal D}\psi] e^{-S}\approx \int
[{\cal D}A ][{\cal D}\psi]{\rm
exp}\left[{-\frac{1}{\Lambda}\int_{M_4} (\delta_{ab} -
\frac{1}{\Lambda}\psi_{ab} + \frac{1}{\Lambda^2}\psi_{ac}\psi_{cb}
+...)F^a\wedge F^b}\right].  \lb{11} \ee
Taking into account that
\be F^a\wedge F^b = \frac 14 F_{\mu\nu}^a F_{\rho\sigma}^b
dx^{\mu}\wedge dx^{\nu}\wedge dx^{\rho}\wedge dx^{\sigma} = \frac
14 F_{\mu\nu}^a F_{\rho\sigma}^b
\epsilon^{\mu\nu\rho\sigma}\sqrt{g}d^4x,     \lb{11a} \ee
we have:
\be F^a\wedge F^b = = \frac 12 F_{\mu\nu}^a
F^{*b\mu\nu}\sqrt{g}d^4x, \lb{11aa} \ee
where
\be   F^{*b\mu\nu} = \frac 12
\epsilon^{\mu\nu\rho\sigma}F_{\rho\sigma}^b  \lb{11b} \ee
is a dual tensor. The requirement of self-duality
$F^{*b\mu\nu}=F^{b\mu\nu}$ gives:
\be   F^a\wedge F^b =  \frac 12 F_{\mu\nu}^a
F^{b\mu\nu}\sqrt{g}d^4x. \lb{11c} \ee
Then the partition function is:
$$
 Z = \int[{\cal D}A ][{\cal D}\psi] e^{-S}\approx \int
[{\cal D}A ][{\cal D}\psi]{\rm
exp}\left[-\frac{1}{2\Lambda}\int_{M_4} (\delta_{ab} -
\frac{1}{\Lambda}\psi_{ab} + \frac{1}{\Lambda^2}\psi_{ac}\psi_{cb}
+...)F^a\cdot F^b\sqrt{g}d^4x \right]  $$ \be = \int[{\cal
D}A]{\rm exp} \left[-{\frac 12}\int_{M_4}
\left(\frac{F^2}{\Lambda} + {\rm ln}(\frac{F^2}{2\Lambda
M^4})\right)\sqrt{g}d^4x \right], \lb{11d} \ee
where  $F^a\cdot F^b \equiv F_{\mu\nu}^a F^{b\mu\nu}$ and $F^2 =
F_{\mu\nu}^a F^{a\mu\nu}$; $M$ is the energy scale parameter .

Assuming the existence of the background $B_{\mu}^a$ of the
connection $A_{\mu}^a$:
\be   A_{\mu}^a = B_{\mu}^a + {\cal A}_{\mu}^a,  \lb{12} \ee
where ${\cal A}_{\mu}^a$ is the perturbative connection, and
$B^a\cdot B^a \equiv B_{\mu}^aB^{a\mu}=\rm{const}$, we can
calculate the partition function (\ref{11d}) and obtain the
effective Lagrangian of the system:
\be Z = \int[{\cal D}{\cal A}]{\rm exp}\left(- \int
L_{eff}\sqrt{g}d^4x\right). \lb{13} \ee
Now we can return to the Minkowski spacetime. Using the gauge
condition ${\cal A}\cdot B = 0$, it is easy to get the following
result:
\be - L_{eff} = \frac1{2\Lambda} \left({\cal F}^a\cdot {\cal F}^a
+ 2\sqrt{F_0^2} {\cal A}^a\cdot {\cal A}^a + F_0^2\right) + \frac
12{\rm \ln}(\frac{F_0^2}{2\Lambda M^4}), \lb{14} \ee
where ${\cal F}^a\cdot {\cal F}^a \equiv {\cal F}_{\mu\nu}^a{\cal
F}^{a\mu\nu}$,  ${\cal F}^a_{\mu\nu} =
\partial_{\mu}{\cal A}^a_{\nu} -
\partial_{\nu}{\cal A}^a_{\mu} + f^{abc}{\cal A}^b_{\mu}{\cal
A}^c_{\nu}$, and ${\cal A}^a\cdot {\cal A}^a\equiv {\cal
A}_{\mu}^a{\cal A}^{a\mu}$.

In Eq.~(\ref{14}) we have:
\be F_0^2 = (B^a\cdot B^a)^2=\rm{const}, \lb{14a} \ee
which means that the vacuum expectation $\langle 0|B^a\cdot
B^a|0\rangle$ is nonzero (this condition is analogous to the gluon
condensate in QCD). We neglected the term $({\cal A}^a\cdot {\cal
A}^a)^2$ in the Lagrangian (\ref{14}), since ${\cal A}_{\mu}^a$ is
a small field perturbation. In general, it is possible to take
into account this quartic term as well.

Choosing $M^4=\rho_{vac}$ and the condition:
\be   F_0^2 = 2\Lambda \rho_{vac}, \lb{15} \ee
we obtain the following physically reasonable result:
\be  - L_{eff} = \frac1{4g^2}[{\cal F}^a\cdot {\cal F}^a
 + m^2 {\cal A}^a\cdot {\cal A}^a] + \rho_{vac},
\lb{16} \ee
when the minimal effective potential density is equal to the
vacuum density of the Universe: $min\,\,U_{eff} =
 \rho_{vac}$.

In Eq.~(\ref{16}) the parameter $m$ is related with a graviton
mass:
\be      m^2 = 2\sqrt{F_0^2} = 2\sqrt{2\Lambda\rho_{vac}} =
\frac{2}{\sqrt3}\rho_{vac}(M_{Pl}^{red.})^{-2}.  \lb{17} \ee
Taking into account an estimate of cosmological measurements for
the vacuum energy density $\rho_{vac}$ of the Universe
\ct{6a,6b,6c,7a,7b,7c}:
\be  \rho_{vac}\approx (2\times 10^{-3}\,\,{\mbox{eV}})^4,
\lb{18} \ee
and the well-known value $M_{Pl}^{red.}\approx 2.43\times 10^{18}$
GeV, we can estimate the mass $m$:
\be
     m \approx (\frac 43)^{1/4}(2\times 10^{-3}\,\,{\mbox{eV}})^2(2.43\times
     10^{18}\,\,{\mbox{GeV}})^{-1}\approx 1.8\times 10^{-42}
     \,\,{\mbox{GeV}}.
\lb{19} \ee
This tiny value of the graviton mass $m$ almost coincides with the
present Hubble parameter value $H_0\approx 1.5\times
10^{-42}\,\,{\mbox{GeV}} $ (see \ct{6a,6b,6c,7a,7b,7c}).

Let us now estimate the effective coupling constant of the
interaction of perturbative gravitational fields. From
Eqs.~(\ref{14})-(\ref{16}) we see that the dimensionless coupling
constant $g$ is:
\be g^2=\frac{\Lambda}{2},  \lb{20} \ee
and according to Eq.~(\ref{10}), we have:
\be  g = \sqrt{(\rho_{vac}/12)}(M_{Pl}^{red.})^{-2}\sim 10^{-61}.
\lb{21} \ee
We see that the perturbative gravitational interaction is
extremely small. Nevertheless, the existence of a tiny graviton
mass $m_g$ leads to the renormalizability of the quantum gravity.

\section{Massive graviton solution}

Could a graviton be massive? The answer to this question seems to
be positive \ct{8a,8b,8c,8d,9,10,11,12,13a,13b,14,15}: if the
graviton Compton wavelength, $\lambda_g = {m_g}^{-1}$, is large
enough ($\sim$ the present Hubble size $H_0$, see the
Eq.~(\ref{19}) of this paper), we should not be able to
distinguish a massive graviton from a massless one. Astrophysical
bounds are even milder \ct{8a,8b,8c,8d}.

A massive graviton in four-dimensions has five physical degrees of
freedom (helicities $\pm 2$, $\pm 1$, 0) while the massless
graviton has only two (helicities $\pm 2$). In the limit $m_g \to
0$, the exchange by the three extra degrees of freedom can be
interpreted as an additional contribution due to one massless
vector particle with two degrees of freedom ("graviphoton"\, with
helicities $\pm 1$) and plus one real scalar particle
("graviscalar"\, with the helicity 0). The graviphotons do not
contribute to the one-particle exchange: the contribution to the
conserved energy-momentum tensor from graviphoton derivative
coupling vanishes. The graviscalar is coupled to the trace of the
energy-momentum tensor and its contribution is generically
nonzero. This is what causes the difference between the theories
of massless and massive gravitons.

The arguments of Refs.~\ct{9,10,11,12,13a,13b,14}, based on the
lowest tree-level approximation  to the calculation of
interactions between two gravitational sources, has a clear
physical interpretation.

In the general case, a theory of massive gravitons possesses
"ghosts"\,. However, if the Lagrangian contains Pauli-Fierz mass
terms \ct{16}, these "ghosts"\, are absent. But in the limit of
vanishing graviton mass ($m_g \to 0$), the graviton propagator
exhibits the Van Dam-Veltman-Zakharov (VDVZ) discontinuity
\ct{10,11,12} originating from the graviscalar which does not
decouple in the massless limit $m_g \to 0$. At the classical
level, the graviscalar doesn't cause problems \ct {12,13a,13b}.
But at the quantum level the theory becomes strongly coupled \ct
{17} (shows nonperturbative effects) at energy scale $ (m^4M _
{Pl}) ^ {1/5}$. This result was confirmed by explicit calculations
in Ref.~\ct{18}. The phenomenon repeats in brane-world models when
gravity is modified in the infrared limit \ct
{19a,19b,20,21,22a,22b}.

Moreover, the nonlinear four-dimensional theory of the massive
graviton is not defined unambiguously. For the vanishing graviton
mass the lowest tree-level approximation breaks down, but the
higher order corrections are singular in the graviton mass. The
next-to-leading terms in the corresponding expansion are huge
since they are inversely proportional to powers of $m_g$. Thus,
the truncation of the perturbative series does not make much sense
and all higher order terms in the solution of classical equations
for the graviton field should be summed up. The summation leads to
the nonperturbative solution which is continuous when $m_g \to 0$.
The perturbative discontinuity shows up only at large distances
where higher order terms are small. These distances are growing
when $m_g \to 0$. In other words, the continuity is not
perturbative and does not depend on distances considered in
theory.

The reason for the problem of the lowest tree-level approximation
is simple: it does not take into account the characteristic
physical scale existing in theory. The nonperturbative calculation
of the Schwarzschild solution leads to the elimination of this
problem \ct{12,13a,13b}. In the nonperturbative approach, the
coupling of the extra scalar mode to the matter is suppressed by
the ratio of graviton mass to the characteristic physical scale
which is Vainshtein's radius \ct{12}: $R_V = (m_g^{-4}R_S)^{1/5}$,
where $R_S$ is the Schwarzschild's radius \ct{13a,13b}. The size
$R_V $ is large for small values of $m_g $. For example, if the
graviton mass is given by Eq.~(\ref {19}), Vainshtein's radius is
larger than Sun radius ($R_V \sim 100$ Kpc). Hence, the
predictions of the massive theory could be made infinitely close
to the predictions of the massless theory due to the tiny $m_g$.

It has been pointed out in Refs.~\ct{14,23} that graviton mass
terms violate Lorentz invariance.

In Plebanski formulation the graviton field $h^{ab}$ is described
as a perturbation \ct{2}:
\be   \Sigma^a = \Sigma^a_0 + \delta \Sigma^a
       = \Sigma^a_0 +  h^{ab}{\bar {\Sigma}}^b_0,  \lb{22} \ee
where the background two-forms are given by:
\be   \Sigma^a_0 = idt\wedge dx^a - \frac 12
\epsilon^{abc}dx^b\wedge dx^c, \lb{23} \ee
and the anti-self-dual forms ${\bar {\Sigma}}^a$ are:
\be {\bar \Sigma}^a = i \theta^0 \wedge  \theta^a + \frac 12
\epsilon^{abc}\theta^b \wedge \theta^c.  \lb{24} \ee
Then
\be  {\bar \Sigma}^a_0 = idt\wedge dx^a + \frac 12
\epsilon^{abc}dx^b\wedge dx^c. \lb{25} \ee
The first step is to find the linearized connection $\delta A^a =
{\cal A}^a$ such that:
\be  d\delta \Sigma^a + \epsilon^{abc} {\cal A}^b \wedge \Sigma^c
= 0. \lb{26} \ee
Then we obtain the equation of massive graviton:
\be   \Box h^{ab} - m_g^2 h^{ab} = 0, \lb{27} \ee
where $m_g=m$ is given by the Eq.~(\ref{19}).

In the Minkowski space background, the metric $g_{\mu\nu}$ can be
expanded as:
\be   g_{\mu\nu} = \eta_{\mu\nu} + h_{\mu\nu},
\lb{28} \ee
where $\eta_{\mu\nu}$ is the flat metric of the Minkowski space.
This expansion was first suggested by R.P.~Feynman \ct{1a}. Then
the equation of massive graviton is:
\be   \Box h^{\mu\nu} - m_g^2 h^{\mu\nu} = 0. \lb{29} \ee

\section{Multi-gravitons and renormalizable quantum gravity}

To make the higher-derivative propagator, the author of
Ref.~\ct{15} explores the model of multi-gravitons. He proposes
the existence of ghost partners for gravitons.

Starting with the Einstein-Hilbert action:
\be L = \frac{1}{16\pi G} R + L_{mat}, \lb{30} \ee
where $L_{mat}$ is the matter Lagrangian, we have the Einstein
equation:
\be G_{\mu\nu} = 8\pi G T_{\mu\nu}, \lb{31} \ee
where $T_{\mu\nu}$ is the energy-momentum tensor.

The gravitons are denoted by $h^{(n)}_{\mu\nu}$, where $n$ runs
from zero to $N - 1$. The real gravitational field is:
\be h_{\mu\nu} = \sum_nh_{\mu\nu}^{(n)}. \lb{32} \ee
All the gravitons are assumed to be massive with Pauli-Fierz mass
terms \ct{16}:
\be   L_{mas} = - \frac{1}{32\pi G\sqrt{-g}}\sum_{n=0}^{N-1}(-1)^n
m_n^2(h_{\mu\nu}^{(n)}h^{(n)\mu\nu} - h^{(n)}h^{(n)}), \lb{33} \ee
where $h=\eta_{\mu\nu}h^{\mu\nu}$, and:
\be
     \sqrt{-g} = 1 + \frac 12 h - \frac 14 h^{\mu}_{\nu}h^{\nu}_{\mu}
     + \frac 18 hh + ...          \lb{34} \ee
Considering $L_{ghost}$ and $L_{kinetic}$, the author of
Ref.~\ct{15})  obtains the following equation in the case $N=1$:
\be
   (\Box - m_0^2) h^{(0)}_{\mu\nu} = - 16\pi
   G\left(T_{\mu\nu} + \frac 13(\frac{\partial_{\mu}\partial_{\nu}}{m^2_0} - \eta_{\mu\nu})T
   \right).
  \lb{35} \ee
It is easy to write the equations of motion in a higher-derivative
theory. For $N=2$ we have:
\be
   (\Box - m_0^2)(\Box - m_1^2) h_{\mu\nu} = (m_0^2 - m_1^2)16\pi
   G\left(T_{\mu\nu} + \frac 13(\frac{\partial_{\mu}\partial_{\nu}}{m^2_0} - \eta_{\mu\nu})T
   \right).
  \lb{36} \ee
Choosing the transverse-traceless gauge (TT-gauge) for all
gravitational fields, we can quantize gravitational wave of each
graviton field, which was performed in Ref.~\ct{15}.

Let us calculate propagators for small $N$. They help to calculate
the vacuum energy and graviton mass corrections. According to the
Feynman rules, the propagator for $N = 2$ is:
\be D_{\mu\nu\rho\sigma} = - i \frac{(m_1^2 - m_0^2)}{(p^2 +
m_0^2)(p^2 + m_1^2)} P_{\mu\nu\rho\sigma},
 \lb{37} \ee
where
\be P_{\mu\nu\rho\sigma} = \frac 12(\eta_{\mu\rho}\eta_{\nu\sigma}
+ \eta_{\mu\sigma}\eta_{\nu\rho} - \frac
23\eta_{\mu\nu}\eta_{\rho\sigma}). \lb{38} \ee
In this case the theory of gravity is renormalizable: the number
of counter terms are finite.

If we wish to avoid the finetuning, then the case $N = 2$ is not
enough. To make a super-renormalizable model \ct{15}, the number
of counter terms are reduced, we must use $N = 4$ with the
assumption: $m_3 = \sqrt{m_0^2 - m_1^2 + m_2^2}$. Then
\be D_{\mu\nu\rho\sigma} = - i \frac{(m_1^2 - m_0^2)(m_1^2 -
m_2^2)(m_0^2 + m_2^2 + 2p^2)} {(p^2 + m_0^2)(p^2 + m_1^2)(p^2 +
m_2^2)(p^2 + m_3^2)} P_{\mu\nu\rho\sigma}.
 \lb{39} \ee
In the same way, we can get a super-renormalizable model for any
even $N$.

Theory \ct{15} preserves unitarity. The higher-derivative
propagators suppress divergences of the vacuum energy and graviton
mass corrections. Applying ghost partners for the Standard Model
particles, quantum gravity with matter fields becomes
renormalizable with power counting arguments.

\section{Conclusions}

In the present investigation we have used the formulation of
general relativity (GR) in terms of self-dual two-forms due to
Plebanski \ct{1}. Using the formulations given in
Refs.~\ct{1,2,3}, we demonstrated that the Plebanski theory
presents quite an economical alternative to the usual metric and
frame-based schemes of GR. In Section 3 we have developed the
nonperturbative theory of gravity with a nonzero background
connection $B_{\mu}^a$: $A_{\mu}^a = B_{\mu}^a + {\cal
A}_{\mu}^a$, where ${\cal A}_{\mu}^a$ is a small perturbative
connection and $a=1,2,3$ is the $SU(2)$ Lie algebra index. We have
assumed that $<0|B_{\mu}^aB^{a\mu}|0>=\rm{const}$. This condition
is analogous to the gluon condensate in QCD. Our prediction gives
a tiny value of the graviton mass: $m_g\approx 1.8\times
10^{-42}\,{\mbox{GeV}}$ which almost coincides with the present
Hubble parameter value $H_0$ \ct{6a,6b,6c,7a,7b,7c}. In this case,
we should not yet be able to distinguish a massive graviton from a
massless one \ct{12,13a,13b}. We also predicted an extremely small
dimensionless coupling constant of the interaction of perturbative
gravitational fields: $g\sim 10^{-61}$.

Considering the problem of renormalizability of quantum gravity,
we briefly reviewed the model of multi-gravitons by H.~Isimori
\ct{15} with $N$ gravitons having masses $m_0,\, m_1,\,...,
m_{N-1}$. For $N=2$, gravity is renormalizable and the number of
counter terms is finite. If $N=4$, then the quantum theory of
gravity is super-renormalizable (the number of counter terms is
reduced), and such a theory avoids any finetuning problems.

We hope to develop the quantum theory of multi-gravitons in our
forthcoming communications.

\section*{Acknowledgements}

I am thankful to Holger Bech Nielsen for fruitful discussions of
the main problems of quantum gravity.


\vspace{3cm}

\begin{thebibliography}{99}

\bibitem{1a} R.P.~Feynman, {\it The Character of Physical Law}, (MIT Press,
1967).
\bibitem{1b} S.~Weinberg, {\it Gravitation and Cosmology}, (ed. by John Wiley and Sons,
New York, 1972).
\bibitem{1c}
C.W.~Misner, K.S.~Thorne, J.A.~Wheeler, {\it Gravitation}, (ed. by
W.H.~Freeman, San Francisco, 1973).
\bibitem{1d}
C.D.~Froggatt and H.B.~Nielsen, {\it Origin of Symmetries}, (World
Scientific, Singapore, 1991).
\bibitem{1e}
G.W.~Gibbons and S.W.~Hawking, {\it Euclidean quantum gravity},
(World Scientific, Singapore, 1993).
\bibitem{1f}
Lee Smolin, {\it Three Roads to Quantum Gravity}, (ed. by
Weindelfeld and Nicolson, London, 2000).
\bibitem{1g}
G.E.~Volovik, {\it The Universe in a Helium Droplet}, (Oxford
University Press, 2003).
\bibitem{1h}
C.~Rovelli, {\it Quantum Gravity}, (Cambridge Monographs on
Mathematical Physics, 2004).
\bibitem{1i}
F.M.~Branley and E.~Miller, {\it Gravity Is a Mystery}, (ed. by
Harper Collins Children Books, 2007).
\bibitem{1j}
T.~Thiemann, {\it Modern Canonical Quantum General Relativity},
(Cambridge Monographs on Mathematical Physics, 2007).
\bibitem{1k}
B.G.~Sidharth, {\it Ether, Space-Time and Gravity}, Vol.3, (ed. by
Michael Duffy, Apeiron Press, USA, 2008).
\bibitem{1l}
V.~Mukhanov, S.~Winitzki, {\it Introduction to Quantum Effects in
Gravity}, (Cambridge Monographs on Mathematical Physics, 2008).
\bibitem{1m}
J.~Ambjorn, B.~Durhuus and T.~Jonsson,  {\it Quantum Gravity: From
Theory to Experimental Search}, Lecture Notes in Physics, (ed. by
Domenico J.W. Giulini, Claus Kiefer, Claus Lammerzahl, Springer,
2009).
\bibitem{1n}
J.~Ambjorn, J.~Jurkiewicz and R.~Loll, {\it Quantum gravity as sum
over spacetimes}, Lecture Notes in Physics {\bf 807}, 59-124
(2010), arXiv:0906.3947[gr-qc].
\bibitem{1p}
T.~Padmanabhan, {\it Gravitation: Foundations and Frontiers},
(Cambridge University Press, 2010).
\bibitem{1q}
Laurent Nottale, {\it Scale relativity and fractal space-time},
(World Scientific Press, 2011).
\bibitem{2a}
 L.V.~Laperashvili, D.A.~Ryzhikh, H.B.~Nielsen, DESY-02-188, Nov 2002,
Int.J.Mod.Phys.A {\bf 18},4403 (2003), hep-th/0211224.
\bibitem{2b}
L.V.~Laperashvili, D.A.~Ryzhikh, {\it Phase transition in gauge
theories and the Planck scale physics}, ITEP-24-01, Moscow, Oct
2001, hep-th/0110127.
\bibitem{2c}
L.V.~Laperashvili, Yad. Fiz. {\bf 57}, 501 (1994) [Phys. Atom.
Nucl. {\bf 57}, 471 (1994)].
\bibitem{3a}
S.~Chadha and H.B.~Nielsen, Nucl.Phys. B {\bf 217}, 125 (1983).
\bibitem{3b}
J.~Ellis, M.K.~Gaillalard, D.Y.~Nanopoulos and S.~Rudaz,
Nucl.Phys. B {\bf 176}, 61 (1980).
\bibitem{3c}
A.~Zee, Phys.Rev. D {\bf 25}, 1864 (1982).
\bibitem{3d}
H.B.~Nielsen and I.~Picek, Nucl.Phys.B {\bf 242}, 542 (1984).
\bibitem{3dd}
H.B.~Nielsen and I.~Picek, Phys.Lett.B {\bf 114}, 141 (1982).
\bibitem{3e}
C.P.~Burgess, J.M.~Cline, E.~Filotas, J.~Matias and G.D.~Moore,
JHEP {\bf 0203}, 043 (2002), hep-ph/0201082.
\bibitem{3f}
J.~Chkareuli, Eur.Phys.J.C {\bf 55}, 309 (2008), arXiv:0802.1950
[hep-th].
\bibitem{3ff}
J.~Chkareuli, Phys.Lett.B {\bf 659}, 754 (2008), arXiv:0704.0553
[hep-th].
\bibitem{3g}
J.L. Chkareuli, C.D. Froggatt, H.B. Nielsen, Nucl.Phys.B {\bf
821}, 65 (2009), hep-th/0610186.
\bibitem{3gg}
J.L. Chkareuli, C.D. Froggatt, H.B. Nielsen, Phys.Rev.Lett. {\bf
87}, 091601 (2001), hep-ph/0106036.
\bibitem{3h}
J.L.~Chkareuli, C.D.~Froggatt, J.G.~Jejelava, H.B.~Nielsen,
Nucl.Phys.B {\bf 796}, 211 (2008).
\bibitem{3i}
D.~Colladay and V.A.~Kostelecky, Phys. Rev. D {\bf 58}, 116002
(1998), hep-ph/9809521.
\bibitem{4a}
M.~Chaichian, P.P.~Kulish, K.~Nishijima, A.~Tureanu, Phys.Lett.B
{\bf 604}, 98 (2004), hep-th/0408069.
\bibitem{4b}
M.R.~Douglas and N.A.~Nekrasov, Rev. Mod. Phys. {\bf 73},  977
(2001), hep-th/0106048.
\bibitem{1} J. Plebanski, Journ. Math. Phys. {\bf 18}, 2511 (1977).
\bibitem{2} K.~Krasnov, {\it Plebanski Formulation of General Relativity: A Practical
Introduction}, Gen.Rel.Grav. {\bf 43}, 1 (2011); arXiv:0904.0423
[gr-qc].
\bibitem{3}
Eyo Eyo Ita, III, arXiv:0911.0604 [gr-qc].
\bibitem{4}
R.~Capovilla, T.~Jacobson, J.~Dell, and L.~Mason, Class. Quant.
Grav. {\bf 8}, 41 (1991).
\bibitem{5}
C F.~Tennie and M.N.R.~Wohlfarth, Phys.Rev.D {\bf 82}, 104052
(2010); arXiv:1009.5595 [gr-qc].
\bibitem{6a}
C.R.~Das, L.V.~Laperashvili, A.~Tureanu, Eur.Phys.J.C {\bf 66},
307 (2010); arXiv:0902.4874 [hep-ph].
\bibitem{6b}
C.R.~Das, L.V.~Laperashvili, A.~Tureanu, AIP Conf.Proc. {\bf
1241}, 639 (2010); arXiv:0910.1669 [hep-ph].
\bibitem{6c}
C.R.~Das, L.V.~Laperashvili, A.~Tureanu, Phys.Part.Nucl. {\bf 41},
965 (2010); arXiv:1012.0624 [hep-ph].
\bibitem{7a}
C.R.~Das, L.V.~Laperashvili, H.B.~Nielsen, A.~Tureanu,
arXiv:1105.6286 [hep-ph].
\bibitem{7b}
C.R.~Das, L.V.~Laperashvili, H.B.~Nielsen, A.~Tureanu, {\it Mirror
World and Superstring-Inspired Hidden Sector of the Universe, Dark
Matter and Dark Energy}, to be published in Phys.Rev.D (2011),
arXiv:1101.4558 [hep-ph].
\bibitem{7c}
C.R.~Das, L.V.~Laperashvili, H.B.~Nielsen, A.~Tureanu, Phys.Lett.B
{\bf 696}, 138 (2011), arXiv:1010.2744 [hep-ph].
\bibitem{8a}
A.S.~Goldhaber and M.M.~Nieto, Phys. Rev. D {\bf 9}, 1119 (1974).
\bibitem{8b}
C.M.~Will, Phys. Rev. D {\bf 57}, 2061 (1998); gr-qc/9709011.
\bibitem{8c}
J.~Uzan and F.~Bernardeau, hep-ph/0012011.
\bibitem{8d}
P.~Binetruy and J.~Silk, astro-ph/0007452.
\bibitem{9}
Y.~Iwasaki, Phys. Rev. D {\bf 2}, 2255 (1970).
\bibitem{10}
H.~van~Dam and M.~Veltman, Nucl. Phys. B {\bf 22}, 397 (1970).
\bibitem{11}
V.I.~Zakharov, JETP Lett. {\bf 12}, 312 (1970).
\bibitem{12}
A.I.~Vainshtein, Phys.Lett. B {\bf 39}, 393 (1972).
\bibitem{13a}
C.~Deffayet, G.R.~Dvali, G.~Gabadadze, A.I.~Vainshtein, Phys.Rev.D
{\bf 65}, 044026 (2002), hep-th/0106001.
\bibitem{13b}
E.~Babichev, C.~Deffayet and R.~Ziour, JHEP {\bf 05}, 098 (2009).
\bibitem{14}
V.A.~Rubakov, {\it Lorentz-violating graviton masses: Getting
around ghosts, low strong coupling scale and VDVZ discontinuity,}
hep-th/0407104.
\bibitem{15}
Hajime Isimori, arXiv:1010.5122 [gr-qc].
\bibitem{16}
M.~Fierz and W.~Pauli, Proc. Roy. Soc. Lond. A 173, 211 (1939).
\bibitem{17}
N.~Arkani-Hamed, H.~Georgi and M.D.~Schwartz, Annals Phys. {\bf
305}, 96 (2003); arXiv:hep-th/0210184.
\bibitem{18}
A.~Aubert, Phys. Rev. D {\bf 69}, 087502 (2004),
arXiv:hep-th/0312246.
\bibitem{19a}
C.~Charmousis, R.~Gregory and V.A.~Rubakov, Phys. Rev. D {\bf 62},
067505 (2000), arXiv:hep-th/9912160.
\bibitem{19b}
R.~Gregory, V.A.~Rubakov and S.M.~Sibiryakov, Phys. Rev. Lett.
{\bf 84}, 5928 (2000), arXiv:hep-th/0002072.
\bibitem{20}
I.I.~Kogan, S.~Mouslopoulos, A.~Papazoglou, G.G.~Ross and
J.~Santiago, Nucl. Phys. B {\bf 584}, 313 (2000),
arXiv:hep-ph/9912552.
\bibitem{21}
G.R.~Dvali, G.~Gabadadze and M.~Porrati, Phys. Lett. B {\bf 485},
208 (2000), arXiv:hep-th/0005016.
\bibitem{22a}
G.R.~Dvali and G.~Gabadadze, Phys. Rev. D {\bf 63}, 065007 (2001),
arXiv:hep-th/0008054.
\bibitem{22b}
G.R.~ Dvali, G.~Gabadadze, X.R.~Hou and E.~Sefusatti,
arXiv:hep-th/0111266.
\bibitem{23}
N.~Arkani-Hamed, H.C.~Cheng, M.A.~Luty and S.~Mukohyama, JHEP {\bf
0405}, 074 (2004), hep-th/0312099.


\end{thebibliography}
\end{document}